# A Fast Unsupervised Scheme for Polygonal Approximation

**BIMAL KUMAR RAY**
School of Computer Science Engineering and Information Systems
Vellore Institute of Technology, Vellore – 632014, INDIA

**ABSTRACT** This paper proposes a fast and unsupervised scheme for the polygonal approximation of a closed digital curve. It is demonstrated that the approximation scheme is faster than state-of-the-art approximation and is competitive with Rosin's measure and aesthetic aspects. The scheme comprises of three phases: initial segmentation, iterative vertex insertion, iterative merging, and vertex adjustment. The initial segmentation is used to detect sharp turns, that is, vertices that seemingly have high curvature. It is likely that some of the important vertices with low curvature might have been missed in the first phase; therefore, iterative vertex insertion is used to add vertices in a region where the curvature changes slowly but steadily. The initial phase may pick up some undesirable vertices, and thus merging is used to eliminate redundant vertices. Finally, vertex adjustment was used to enhance the aesthetic appearance of the approximation. The quality of the approximations was measured using the Rosin's method. The robustness of the proposed scheme with respect to geometric transformation was observed.

## I INTRODUCTION

The boundary of a two-dimensional digital object can be represented by the Freeman eight-direction chain code. Most 2D digital objects have a large number of pixels (points) at their boundaries. The storage requirements for these curves are high, but many digital points are not necessary to describe the intrinsic shape of an object; rather, a few of them are necessary and sufficient to recognize a shape using a boundary-based shape representation, analysis, and recognition scheme. One of the schemes that can eliminate undesirable points from a digital boundary while retaining its intrinsic shape is its polygonal approximation, which represents a digital boundary by a sequence of straight line segments that are obtained by joining the salient points on the boundary, called the vertices of a polygonal approximation. There exists a polygonal approximation technique that not only represents a digital boundary with the least possible number of vertices, but also produces a visual aesthetic approximation. Decades of research on the problem of representing a digital boundary by a polygon has resulted in a large number of schemes for polygonal approximation, but the problem is yet to be solved.

There are supervised and unsupervised schemes for polygonal approximation. Supervised schemes require user intervention to specify either the number of vertices or the maximum allowable error (called *the threshold*) on the maximum error of approximation. Various schemes are supervised in nature, namely, iterative splitting, iterative split-and-merge, sequential scanning, and iterative point elimination. Specifying either the number of vertices or the approximation error in a supervised scheme is a fundamental scale problem. The choice of either a small number of vertices or, equivalently, a large approximation error results in a distorted approximation that fails to retain the intrinsic shape of a curve. In contrast, selecting either a large number of vertices or equivalently a low threshold value for the approximation error leads to an approximation with too many vertices, many of which are redundant for retaining the intrinsic shape of a curve. Although supervised schemes suffer from the problem of scale, they have the advantage of being able to approximate a digital curve using a polygon with any desired degree of accuracy. If it is necessary to analyze a curve at various levels of detail, a supervised scheme can be used. There also exists optimal supervised scheme that determines the approximation either minimizing the sum of square of errors for a specified number of vertices (known as $min - \epsilon$ problem) or minimizing the number of vertices with a specified approximation error (known as $min - \#$ problem). By contrast, an unsupervised scheme does not require user intervention and can generate a polygonal approximation based on the intrinsic geometric characteristics of an input curve.

A fast and unsupervised scheme for the polygonal approximation of a closed digital curve was proposed in this study. It is observed that the approximation is aesthetic, has appreciable merit with respect to Rosin's measure, is faster than the state-of-the-art scheme, and in terms of robustness to geometric transformation. An overview of the various schemes for polygonal approximation is presented in the next section, the metrics that are frequently used to assess the quality of an approximation are outlined in Section III, the methodology for polygonal approximation reported in this communication is presented in Section IV, the experiments and their analysis are presented in Section V, and finally, in Section VI, the conclusion is drawn and future research direction for the problem is mentioned.

## II. AN OVERVIEW

Several studies addressing the problem of polygonal approximation have been published. A good solution to this problem can be used in tasks such as object identification, character recognition, and signature verification, in addition to the other problems of computer vision, image processing, and pattern recognition. There are various approaches to this problem and heuristic solutions dominate the literature. A detailed review of some of these approaches can be found in [23]. The oldest solution to the polygonal approximation problem is derived from the convex hull determination technique. The iterative splitting technique of Ramer [1] and Douglas and Peucker [2] is similar to the convex hull algorithm. The deficiencies of the scheme were later alleviated by other researchers after incorporating symmetry [3] into the approximation, adjusting vertices to improve approximation error by reducing the sum of square errors [4], and making the approximation independent of initial segmentation [5]. Iterative splitting was followed by iterative split-and-merge by Pavlidis and Horowitz [6]. The approximations generated can be enhanced by vertex adjustment to make them independent of the initial segmentation and to reduce the approximation error. Following iterative splitting and iterative split-and-merge, the sequential scheme by Williams [9] followed by another sequential but faster scheme by Wall and Danielsson [10].

A paradigm shift from these approaches is iterative point elimination introduced by Pikaz and Dinstein [11] and Visvalingam and Whyatt [12]. The conventional scheme for polygonal approximation looks for the vertices of a polygon, whereas iterative point elimination addresses the complementary problem of looking for points on a curve that are not likely to be the vertices of a polygon approximating a curve. For this reason, Masood [13] named this approach for polygonal approximation as *reverse polygonization*. Following the initiation of iterative point elimination, several authors have contributed to the solution of the problem using different metrics to decide upon the points on a curve that need to be eliminated. Masood also used vertex adjustment [4] and found approximations that were close to the optimal solution, albeit with increased computational load. Carmona-Poato et al. [14] used a sequential technique similar to Williams' scheme to iteratively eliminate points that are not likely to be the vertices of a polygon. The initial phase of the scheme detects the breakpoints of a curve, and in the next phase, points are further deleted using the strongest vertex as the starting vertex of the approximation. The scheme is parametric in nature, defining the parameter as the ratio of the decrease in the length of a curve due to point elimination and the maximum approximation error. The approach, being sequential in nature, has the same drawback as that of conventional sequential techniques, viz. rounding off sharp features. This scheme was converted into an unsupervised scheme [15], using the minimum value of $WE_2$ as the stopping criterion for iterative point elimination. Augilera-Augilera et al. [16] used the concept of a concavity tree to determine an initial segmentation of a curve and showed improved performance in terms of execution time on iterative point elimination schemes such as Masood [13] and Carmona et al. [14]. The segmentation scheme is iterative and non-parametric (unsupervised), and it considers various levels of detail of a curve. It repeatedly identifies the convex hull of the concave segments of a curve that closes an open segment with its cover. The measure $WE_2$ is used for non-parametric iterative decompositions. Fernandez-Gracia et al. [3] determined an initial coarse segmentation of a curve using the furthest point(s) from its centroid and the furthest point(s) from the 'already determined furthest point(s) from the centroid' as the set of vertices for an initial segmentation, and the points obtained are called initial points; then, they used a symmetric version of the Ramer and Doughlus-Pecker iterative splitting scheme to determine non-initial points. The significance level of the non-initial points is assigned a significance value as its absolute perpendicular distance from the line segment joining its neighboring initial points. Because the significance value of the initial points must be greater than the significance value of the non-initial points, the significance value of the initial points is determined based on the maximum significance value of the non-initial points and the maximum distance of the points of a curve from its centroid. If the maximum significance value of the non-initial points is zero, then the significance value of the initial points is taken as unity. A normalized significance curve is generated, and four thresholding alternatives, namely, proximity, distance, Rosin, and adaptive, are proposed to automatically detect the threshold, and hence, a polygonal approximation of a curve. The adaptive thresholiding strategy produces the best result among the four alternatives, except in the singular case, wherein proximity thresholding is recommended. This procedure is non-parametric and produces a symmetric approximation from a symmetric curve. Madrid-Cuevas et al. [5] used convex hull decomposition for an initial segmentation, inserted additional points using the non-parametric framework of Prasad et al. [17], and used a three-point merging strategy through optimization of the $WE_2$ measure. The procedure is non-parametric, and the approximations produced by the scheme are aesthetic. Fernandez-Gracia et al. [18] modified [3] using convex hull decomposition of a curve for initial segmentation. The vertices obtained are called initial points. They then performed iterative splitting of the initial segmentation to determine non-initial points. The significance values of the initial and non-initial points are determined in the same manner as in [3], and a significance curve is generated to derive the threshold using the adaptive thresholding introduced in [3]. The approximation is then subjected to refinement through the elimination of quasilinear vertices followed by vertex adjustment, as in Masood's stabilized scheme [4]. This procedure is unsupervised.

Pervez and Mauhmud [19] proposed a non-parametric scheme that relies on high-curvature points to segment a curve further. The high curvature points, also called cut-points, are determined starting with the break points of a curve and then iteratively removing the break points using constrained collinear point suppression of the point with the smallest support region and adjusting the support region of the neighboring points after the elimination of a point. If multiple points are found to be candidates for elimination because of equal strength, then the distance to the centroid is taken to break the tie. The constrained collinear point suppression is performed iteratively starting with a threshold of 0.5 on the perpendicular distance and incrementing it in a step size of 0.5 until two successive iterations do not result in a different number of vertices. The final value of the threshold was then used to further decompose the segments of a curve defined by two consecutive cut points. Each segment is again subjected to constrained collinear point suppression using its break points as the initial set of vertices and then merging the vertices with increasing values of the threshold in a step size of 0.5 until it reaches its final value. The set of vertices of a segment that results in the minimum value of the weighted figure of merit within the segment is used to approximate the segment. The polygonal approximation of a curve is the union of piecewise linear approximations of the segments. Parvez [20] performed constrained collinear point suppression as in [19] and called the final threshold the neighborhood size of a vertex. He either eliminates a vertex or relocates the same depending on whether the elimination error is greater (or less) than the relocation error restricting the relocation region within the estimated neighborhood of a vertex. The vertex with the smallest strength is a candidate for relocation/deletion. The approximation scheme is non-parametric; however, in contrast to [19], it does not force the vertices to lie on the boundary of a curve.

## III. MEASURE OF QUALITY OF APPROXIMATION

A closed digital curve is defined as a circular sequence of digital points. A polygonal approximation scheme represents a curve using a sequence of straight-line segments. When a curve is approximated by a polygon, the compression ratio of the approximation is defined by the ratio of the number of digital points $(n)$ on a curve to the number of vertices$(m)$, the sum of squares of errors $E_2$ is defined by the sum of the square of the perpendicular distance of the digital points from the approximating line segments; and the maximum error $E_\infty$ is the maximum absolute perpendicular distance of all the digital points from approximating line segments. The higher the compression ratio, the higher the values of the total error and maximum error; therefore, only one of these measures cannot reflect the quality of a polygonal approximation scheme. The earliest of the measures used to assess the quality of a polygonal approximation was introduced by Sarkar [21] and is called the figure of merit. It is defined as the ratio of the compression ratio to the sum of the square of the errors. The researchers later preferred to use its reciprocal and called it the weighted figure of merit. The other forms of weighted figure of merit use the second power of the compression ratio (and this measure is denoted by $WE_2$) and the third power of the compression ratio (denoted by $WE_3$) with the sum of squares of errors in the numerator of the ratio. Another measure for assessing the quality of approximation is the ratio of the maximum error to the compression ratio (denoted by $WE_\infty$).

Rosin [8] criticized Sarkar's measure and introduced a measure defined by the geometric mean of two ratios, fidelity and efficiency, expressed as a percentage. Fidelity is defined by the ratio of the approximation error as obtained from an optimal scheme using the number of vertices returned by a suboptimal scheme to the approximation error produced by a suboptimal scheme, which is defined as the ratio of the number of sides required by an optimal scheme to generate an approximation with approximation error of a suboptimal scheme to the number of sides generated by a suboptimal scheme. This measure will henceforth be called Rosin's measure. The computation of Rosin's measure involves the use of an optimal scheme for polygonal approximation, which results in high computational load. An approximate version of the optimal scheme was used in this study to compute Rosin's measure.

## IV. PROPOSED METHODOLOGY

The proposed methodology comprises three phases: initial segmentation, iterative vertex insertion, iterative merging, and vertex adjustment. The initial segmentation involves the detection of high-curvature points by scanning a curve in clockwise and counterclockwise directions. Starting from an arbitrary point on a curve, it is scanned by a line segment joining the starting and endpoints of the line segment. The length of the line segment increases as the scanning progresses, except when the line passes through a sharp turning point (high curvature point). The length of the scanning line segment decreased after it crossed a sharp turn [10]. Consider three points $p_s$, $p_t$ and $p_e$ $(e > s \; and \; e = t + 1)$ on a curve, as shown in the figure below (Figure 1). The curve point $p_s$ and $p_e$ are the endpoints of a scanning line segment, and the curve point $p_t$ is an intermediate point on the curve. If there is a sharp turning at $p_t$ then the length of segment $\overline{p_s p_e}$ decreases as scanning line $\overrightarrow{p_s p_e}$ passes by $p_t$.

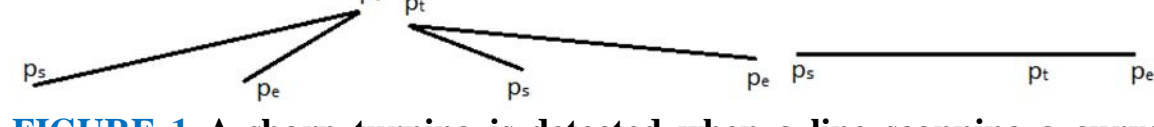

**FIGURE 1** **A sharp turning is detected when a line scanning a curve experiences reduction in its length.**

This observation reveals that as the scanning line segment passes through a high-curvature point, it can be taken as a vertex (at least temporarily) of a polygonal approximation of a curve. A curve is scanned whenever the length of the scanning line segment joining the starting point to the last point $(p_{t+1})$ visited is found to be smaller than the last line segment joining the starting point $p_s$ to the point $p_t$ then $p_k$ can be designated as a probable vertex of a polygon approximating a curve. Thus, the sharp turns of a curve can be located using the observed length of a line segment scanning the curve. However, this may also result in noisy sharp turnings, which can be deleted through a subsequent merging phase (presented later). As a curve is scanned, every time a sharp turning is detected, scanning is initiated from there on, and this process is repeated until a sharp turning detected coincides with one of the sharp turnings already detected. One scan of the entire curve detects one set of sharp turns. Because a clockwise scanning of a curve in general produces a different set of vertices than its counter-clockwise scanning and vice versa, a curve is scanned twice, once in the clockwise direction and once in the counter-clockwise direction, and the union of the vertices obtained in two passes is taken as the initial set of vertices. The computational complexity for detecting vertices through scanning is $O(n)$ because each point of a curve is visited only once in each pass (clockwise and counterclockwise). The detection of sharp turns is not computationally intensive because it involves comparing one Euclidean distance with another; thus, the time to detect sharp turns is low. The following figure (Figure 2) shows the digital curve (in red) and its initial segmentation points (in yellow) overlaid on the digital curve.

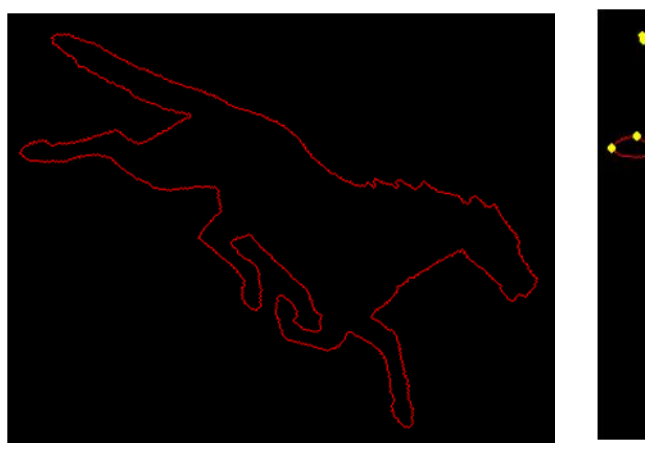
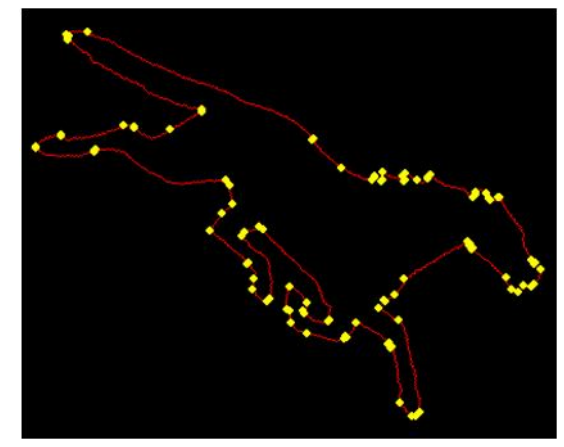

**FIGURE 2** **A digital curve (left) and its initial segmentation (right)**

The above scheme for initial segmentation fails to detect low-curvature points, although its support region may be long enough for the peak/valley point located in the region to be caught by a human visual system. An example of a low-curvature point with a long support region is shown in the following figure (Figure 3).

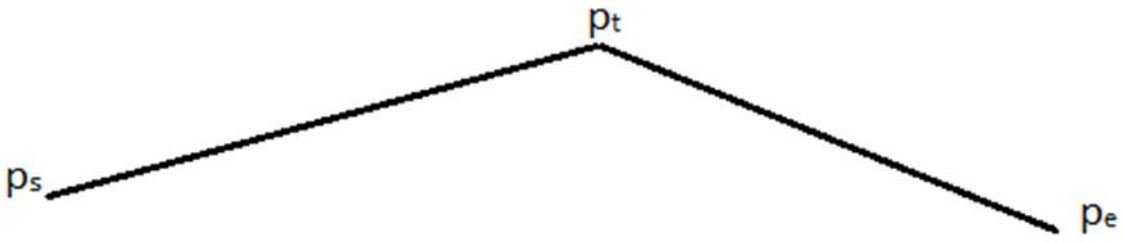

**FIGURE 3** **Significant vertices may be missed in search of sharp turnings only.**

In Figure 3, although there is a turning at $p_t$ but the turning is not detected by the scanning technique described above because the scanning line continues to grow as it passes point $p_t$. However, these vertices can be inserted into an approximation through a subsequent iterative vertex-insertion phase, as described in the following paragraph.

The initial segmentation of a curve defined by sharp turns is subjected to iterative decomposition through vertex insertion, until a certain criterion is satisfied. Let a curve with $n$ digital points have $q$ segments $S_1$, $S_2$, … $S_q$, $S_{q+1}$ determined by its sharp turns, where the segment $S_{q+1} = S_1$ because of the circular nature of a closed curve, and the segment $S_r(1 \leq r \leq q)$ be defined by the consecutive points $p_i$ to $p_{i+l}$ $(l \geq 1)$ of a curve. These segments are subjected to further decomposition in the iterative vertex insertion phase using the maximum error in a segment to determine the candidate segment for decomposition, and inserting a vertex in the segment to minimize the sum of squares of errors over the segment. The maximum error in segment $S_r$ is the maximum absolute perpendicular distance of the points on the segment. The segment with the highest maximum error among all segments is decomposed into two segments such that the sum of the square of errors of the approximation of $S_r$ by two segments is minimized. One vertex was inserted into a segment with the maximum error in the iteration. A heuristic function defined by the sum of the number of vertices of the intermediate approximation and the sum of the square of the errors of the approximation was used as the stopping criterion for decomposition. The rationale in favor of the choice of the heuristic function is that the objective of a polygonal approximation is to reduce the number of vertices as well as the total error of the approximation. An attempt has also been made with the second and third powers of the number of vertices instead of the first, but over-smoothing has been experienced. Decomposition is repeated as the value of the heuristic function decreases. Initially, the function had a very high value because of a few vertices and a large error. Every time a new vertex is inserted, the approximation error decreases significantly, but the number of vertices only increases by unity. An approximation that results in a local minimum value of the heuristic function improves the initial segmentation. The computational complexity of the iterative vertex insertion is $O(n \log n)$. The image on the left in Figure 4 shows the results of the initial segmentation followed by iterative vertex insertion. The vertices are indicated in cyan and are overlaid on the input digital curve.

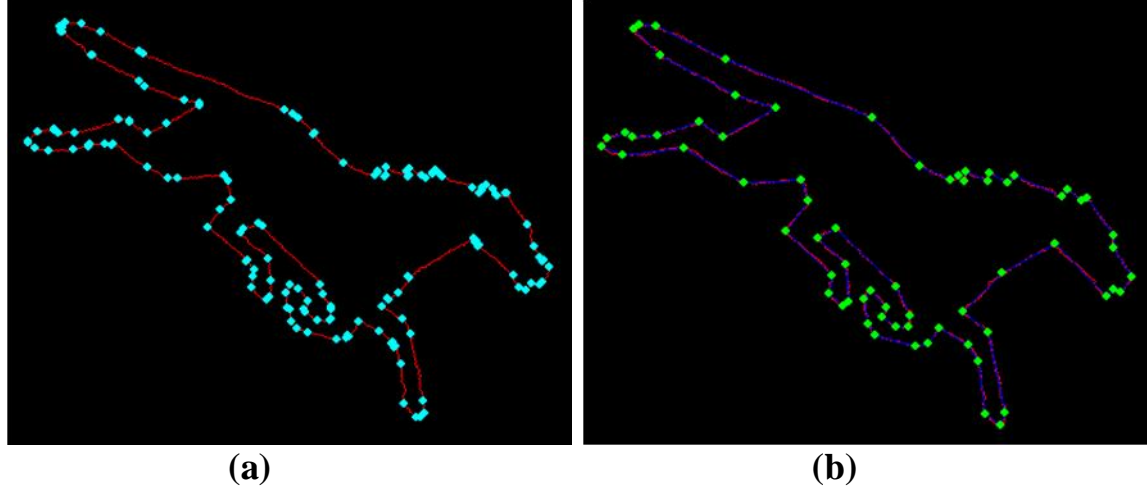

**FIGURE 4 (a) The vertices (in cyan) after initial segmentation followed by iterative vertex insertion overlaid on the digital curve, (b) the polygonal approximation overlaid on the digital curve with vertices in green and sides of the polygon in blue after iterative merging followed by vertex adjustment.**

Some of the vertices obtained through initial segmentation could have been noisy and hence were not necessary to retain the intrinsic shape of a curve; thus, a merging step was adopted following iterative vertex insertion. The merging step is an attempt to remove pseudo-vertices to generate an aesthetic approximation. This phase of the approximation involves the computation of the absolute perpendicular distance (called vertex error) of a vertex from the line segment joining its adjoining vertices. Vertex with the least vertex error is called the weakest vertex. The merging phase attempts to delete the weakest vertex. To ensure that deletion of the weakest vertex does not result in an undesirable approximation, the absolute perpendicular distance of the vertices that are outside the line segment is also computed. If the absolute perpendicular distance of the weakest vertex is less than the maximum approximation error ($\delta$ say, obtained after the iterative vertex insertion phase) and the absolute perpendicular distance of the other vertices (that are outside the segment) from the line segment enclosing the weakest vertex is greater than the maximum approximation error ($\delta$), the vertex is eliminated and the errors of the neighboring vertices are modified. The rationale for considering the additional perpendicular distance of other vertices is that otherwise, deletion of a vertex may lead to a self-intersecting approximation, although the input digital curve is not self-intersecting [19]. Merging is performed iteratively by looking for the weakest vertex until all such vertices have been considered as candidates for merging. The computational complexity of iterative merging is $O(n \log n)$.

After the undesirable vertices are eliminated through iterative merging, the positions of all vertices are adjusted to minimize the sum of the squares of the approximation errors. The vertex adjustment was initiated from the strongest vertex. The strength of a vertex is defined by the sum of the lengths of its adjoining sides, and the vertex with the highest strength is selected as the starting vertex to initiate the vertex adjustment. The rationale in favor of the choice of the starting vertex is the fact that this vertex has the highest flexibility in its positional movement to reduce the sum of squares of errors. Masood's stabilization scheme [4] performs vertex adjustment after the deletion of a vertex, and vertex adjustment is initiated from vertices adjacent to the deleted vertex. The proposed scheme also performs vertex adjustment, but it does so after merging all candidate vertices, and initiates vertex adjustment with the strongest vertex. The process was terminated after all vertices had been tested for adjustment. The computational complexity of vertex adjustment is $O(m'n)$; where $m'$ is the number of vertices obtained at the end of the iterative merging process.

To ensure that merging followed by iterative vertex adjustment does not result in undesirable distortion in the approximation, merging followed by iterative vertex adjustment is continued for a finite number of iterations until the approximation is stabilized. The merging is terminated when two consecutive iterations result in the same value of the weighted figure of merit $WE_\infty$ and $WE_2$ (the measures $WE_\infty$ and $WE_2$ reflect the smoothness of an approximation). The overall computational complexity of the polygonal approximation scheme was $O(m'n)$. The image on the right in Figure 4 shows the polygonal approximation overlaid on the digital curve (depicted in red). The sides of the polygon are indicated by a blue line segment, and the vertices are indicated in green. The methodology described in this section is consolidated into the following algorithm:

**Algorithm: Polygonal approximation**
Input: A closed digital curve
Output: Its polygonal approximation
Begin
Step 1 Determine sharp turns through curve scanning in the clockwise and counterclockwise directions.
Step 2 Insert vertices iteratively as long as a specific heuristic criterion is satisfied.
Step 3: Merge vertices iteratively, ensuring that the approximation does not result in undesirable distortions.
Step 4: Perform iterative vertex adjustment starting from the strongest vertex to minimize the sum of the squares of errors.
Step 5 Repeat Step 4 and 5 until the approximation is stabilized
Step 6 Join the consecutive vertices by straight line segment.
End.

## V. EXPERIMENTS AND ANALYSIS

The MPEG7 dataset [22] was used to validate the quality, efficacy, and robustness of the approximations generated by the proposed methodology. The approximations were also compared with those generated by the scheme of Madrid-Cuevas et al. [5]. Although there are other schemes for polygonal approximation that are unsupervised in nature, the quality of approximation produced by the scheme of Madrid-Cuevas et al. has been found to be better than those generated by other unsupervised schemes. Table I shows the polygonal approximations generated by the proposed scheme and those generated by the Madrid-Cuevas et al. scheme, overlaid on the digital curves from the MPEG7 dataset. The digital boundary is indicated in red, sides of the polygonal approximation in blue, and vertices in green.

**TABLE I. Approximations generated by the proposed scheme and by the Madrid-Cuevas et al. scheme.**

| Digital curve | Approximation by the proposed scheme | Approximation by Madrid-Cuevas et al. scheme |
|---|---|---|
| Apple | | |
| Bat | | |
| Beetle | | |
| Bell | | |
| Bird | | |
| Bone | | |
| Bottle | | |

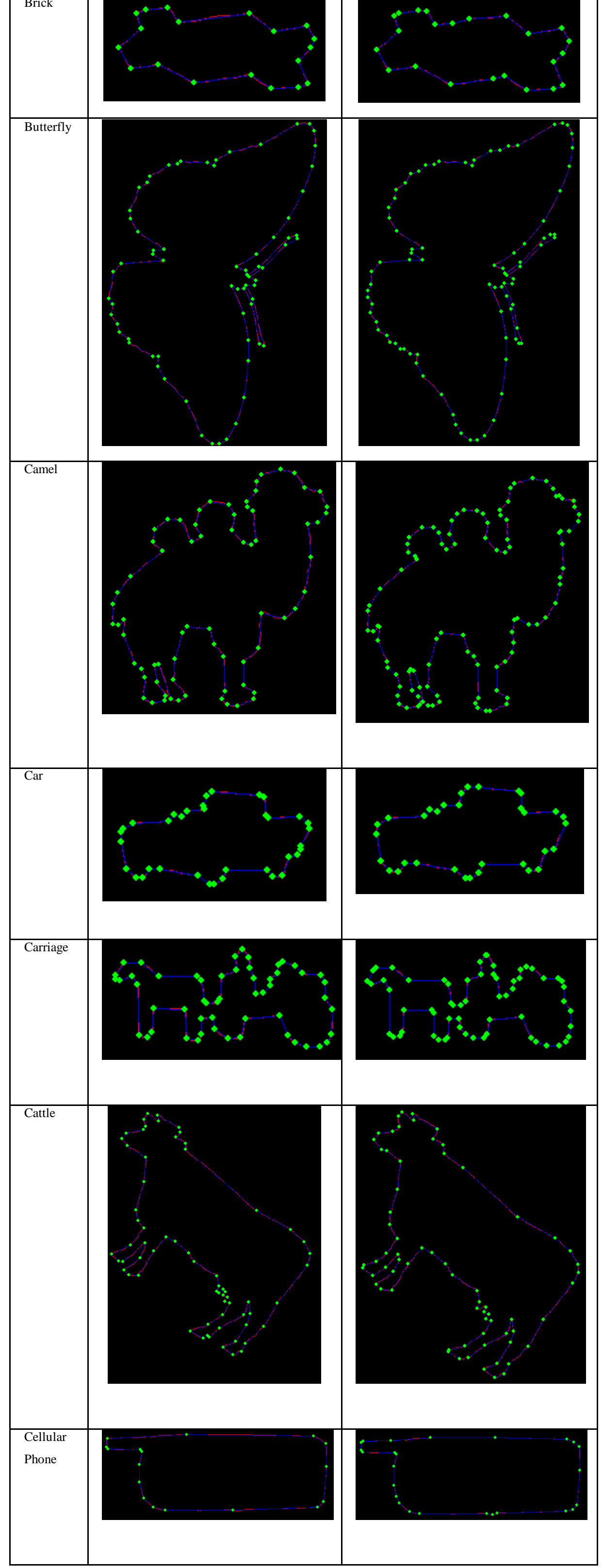

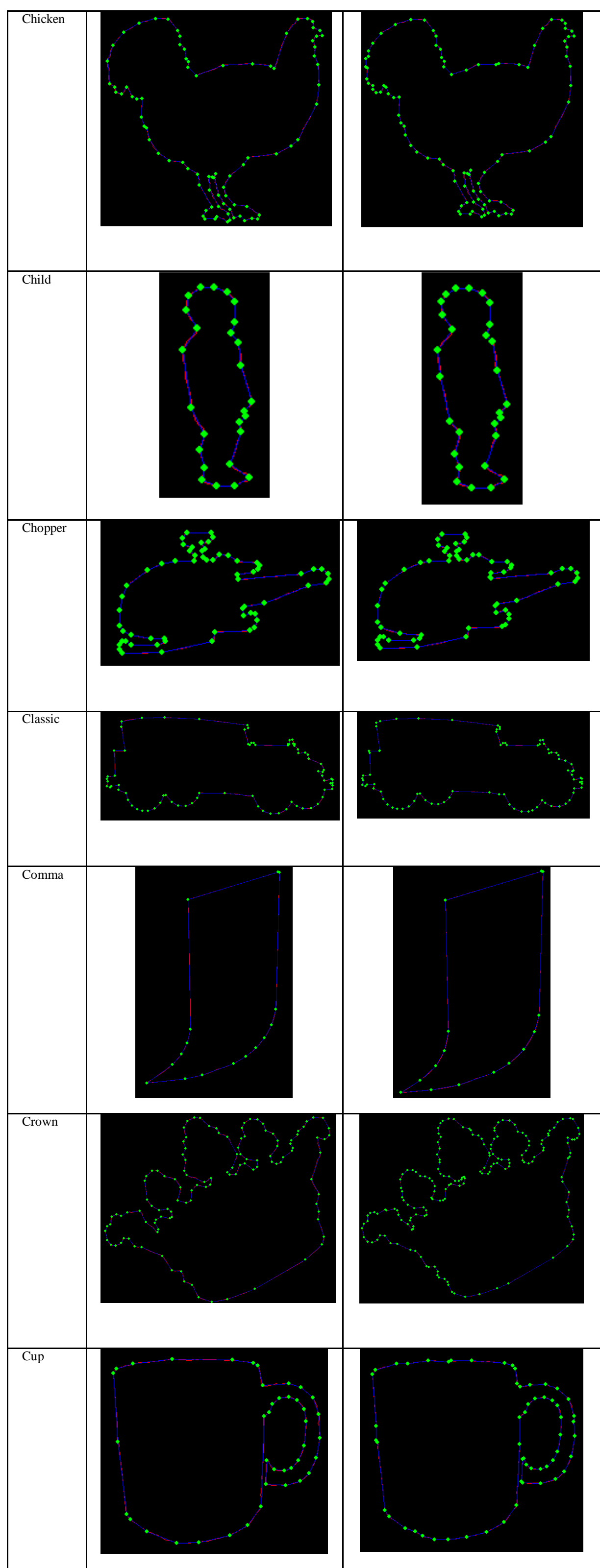

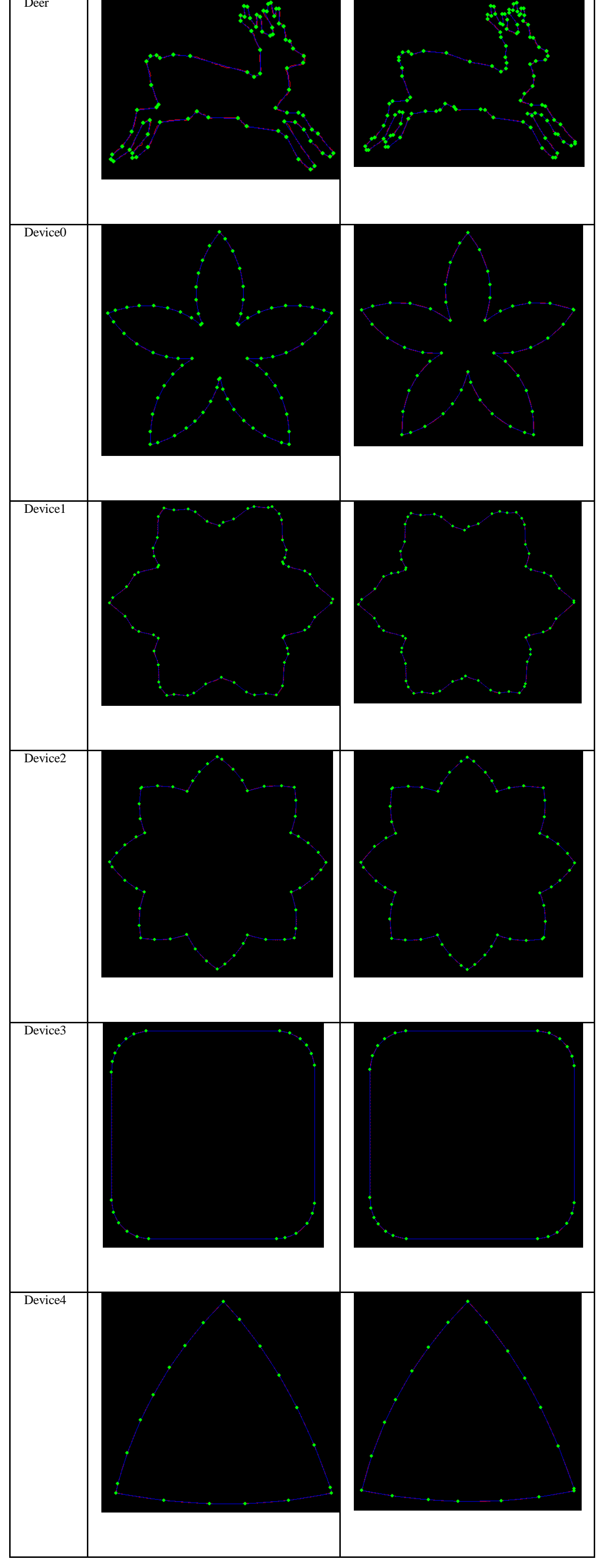

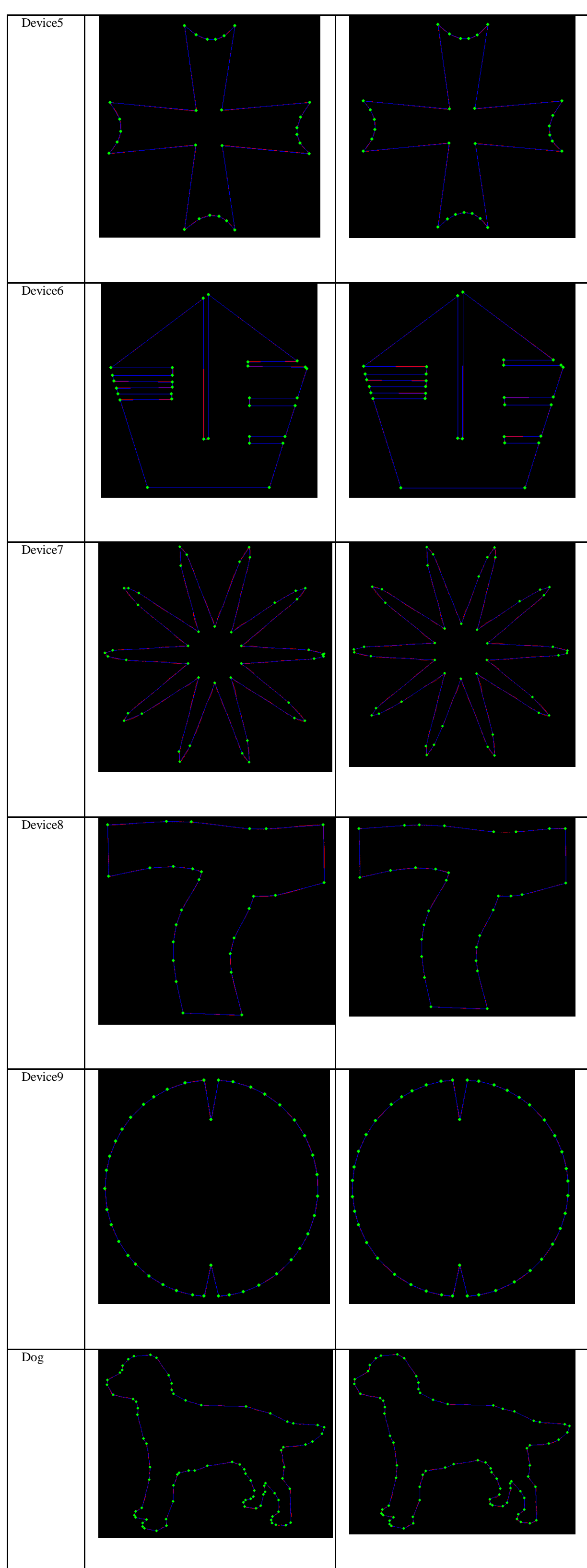

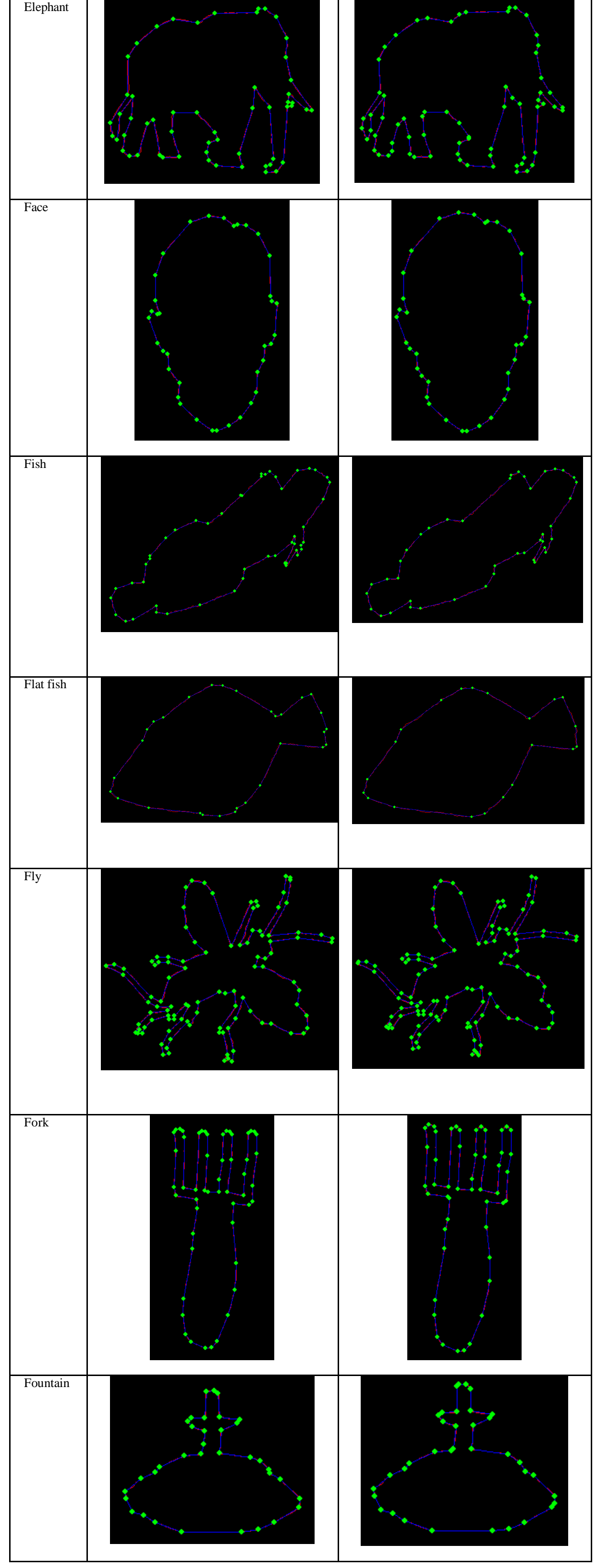

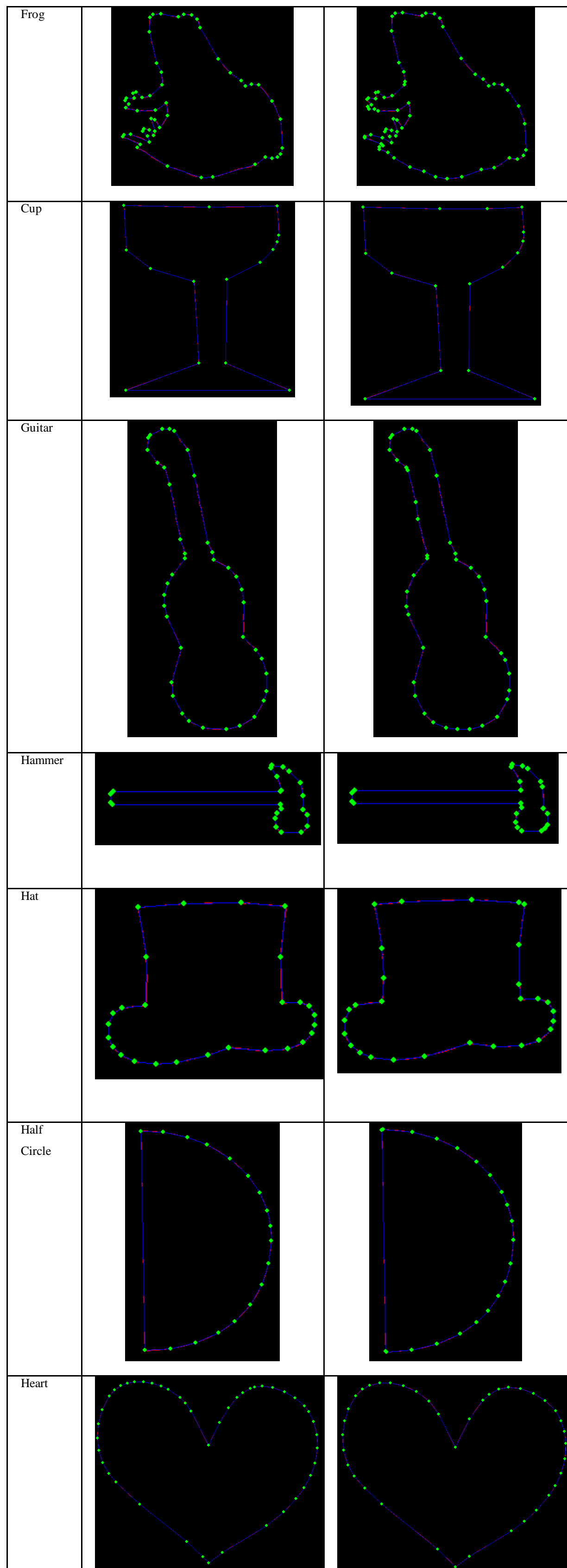

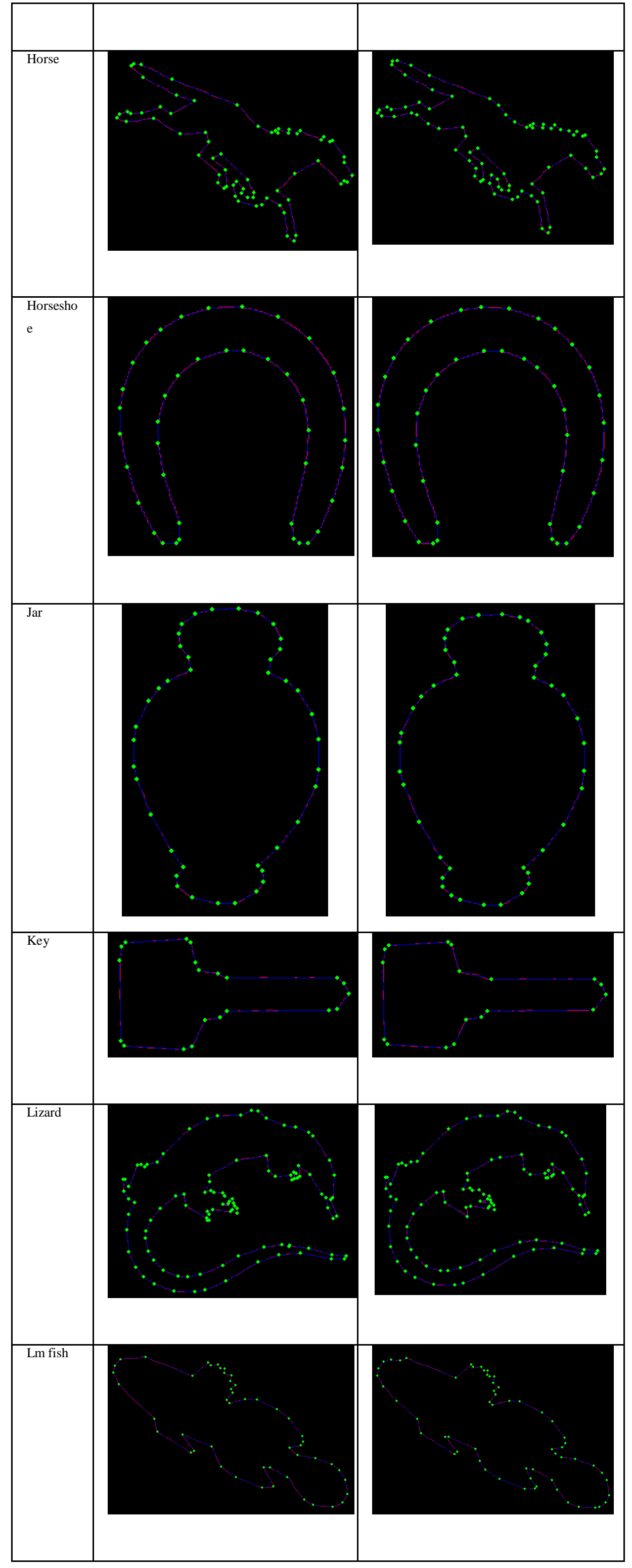

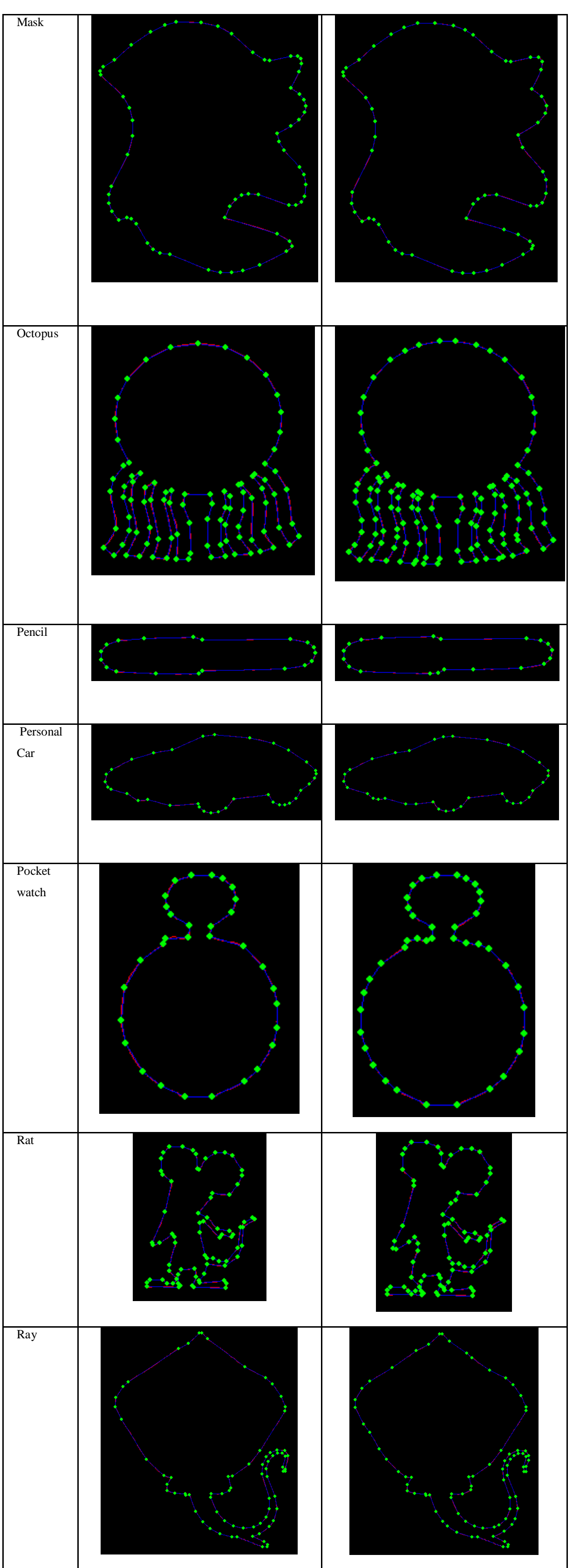

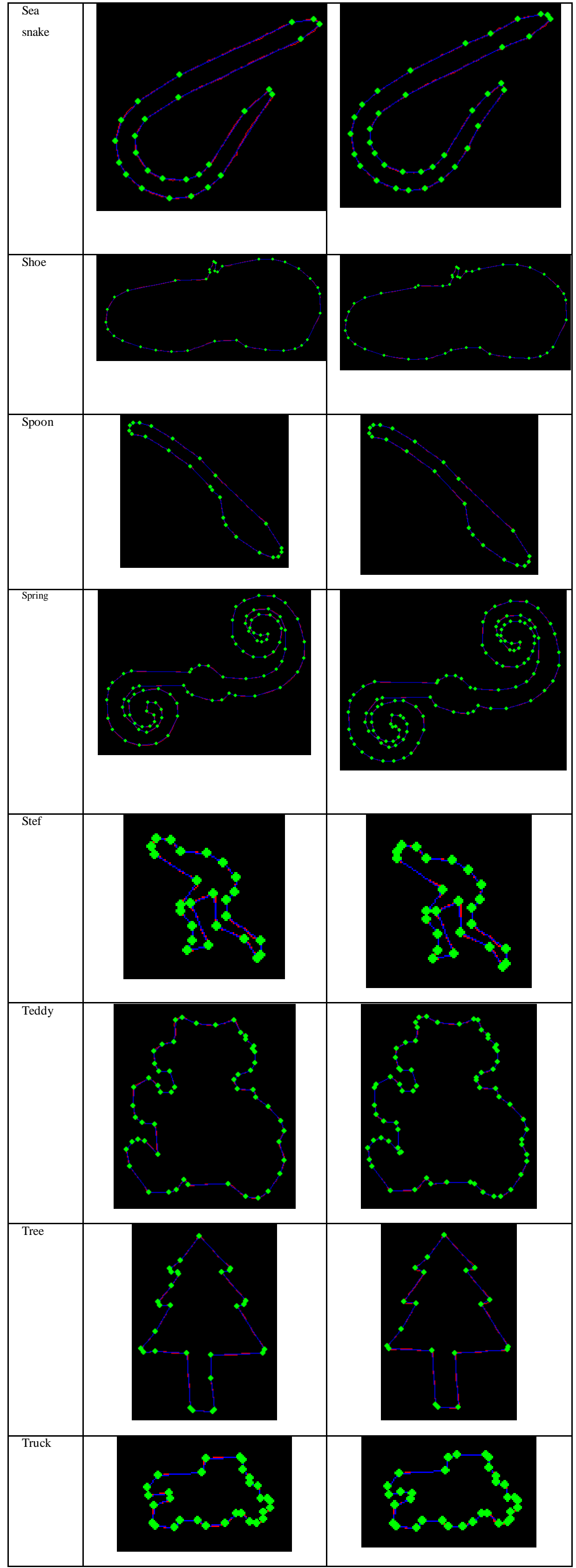

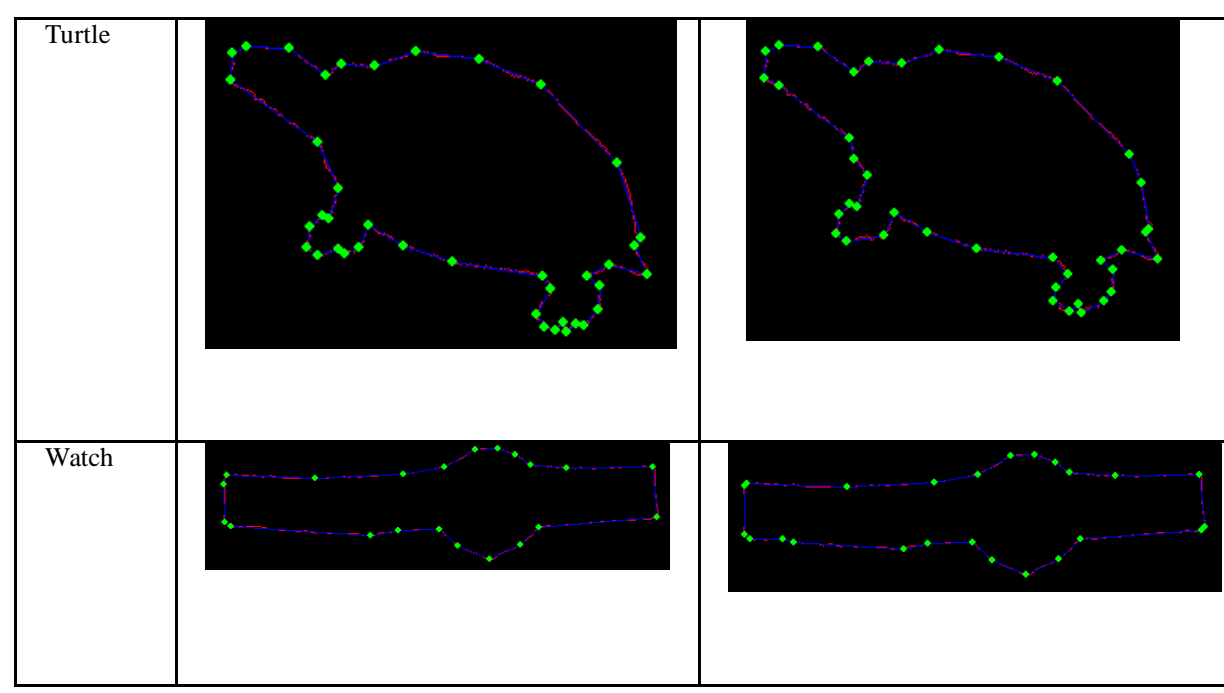

A close look into the approximations generated by the proposed scheme and those generated by the Madrid-Cuevas et al. scheme reveals that the two approximations are almost similar. The approximations produced by these two schemes are competitive in terms of aesthetics. However, the proposed scheme detects sharp turnings more faithfully than the Madrid-Cuevas et al. scheme.

The approximations were also compared using Rosin's measure, as shown in Figure 5. The horizontal axis indicates the digital curves of the MPEG7 dataset used for testing and the vertical axis indicates the value of Rosin's merit measure. The blue line diagram is the plot of the merit measure for the Madrid-Cuevas et al. scheme, and the green line diagram is the plot of the merit measure for the proposed scheme. The performance of the proposed scheme was found to be competitive with that of the Madrid-Cuevas et al. scheme with respect to Rosin's measure. It is worth mentioning that the Madrid-Cuevas et al. scheme, in general, produces more vertices than the proposed scheme, and it is well known that Rosin's measure is biased towards approximations with a large number of vertices. Perez and Vidal's [7] optimal scheme was used to compute Rosin's measure. The computational complexity of Perez and Vidal is significantly high, which is why an approximate version of the scheme [23] is used here.

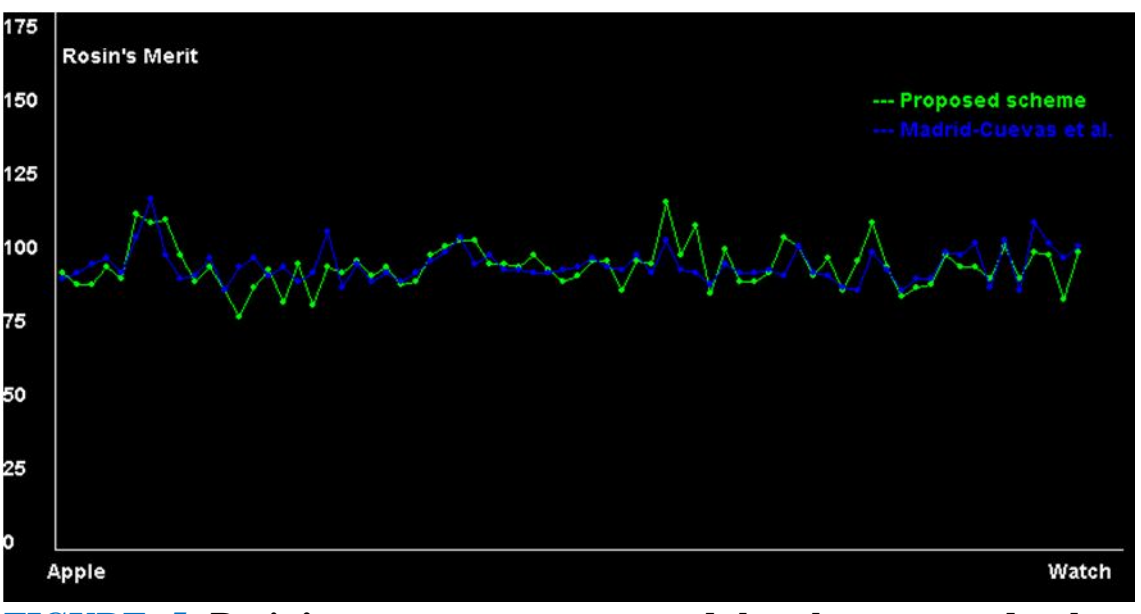

**FIGURE 5 Rosin's measure as generated by the proposed scheme (green line diagram) and by the Madrid-Cuevas et al. scheme (blue line diagram).**

The proposed scheme is compared with the Madrid-Cuevas et al. scheme with respect to execution time, and it is found that the proposed scheme is faster than the Madrid-Cuevas et al. scheme, as depicted in the following line diagram (Figure 6).

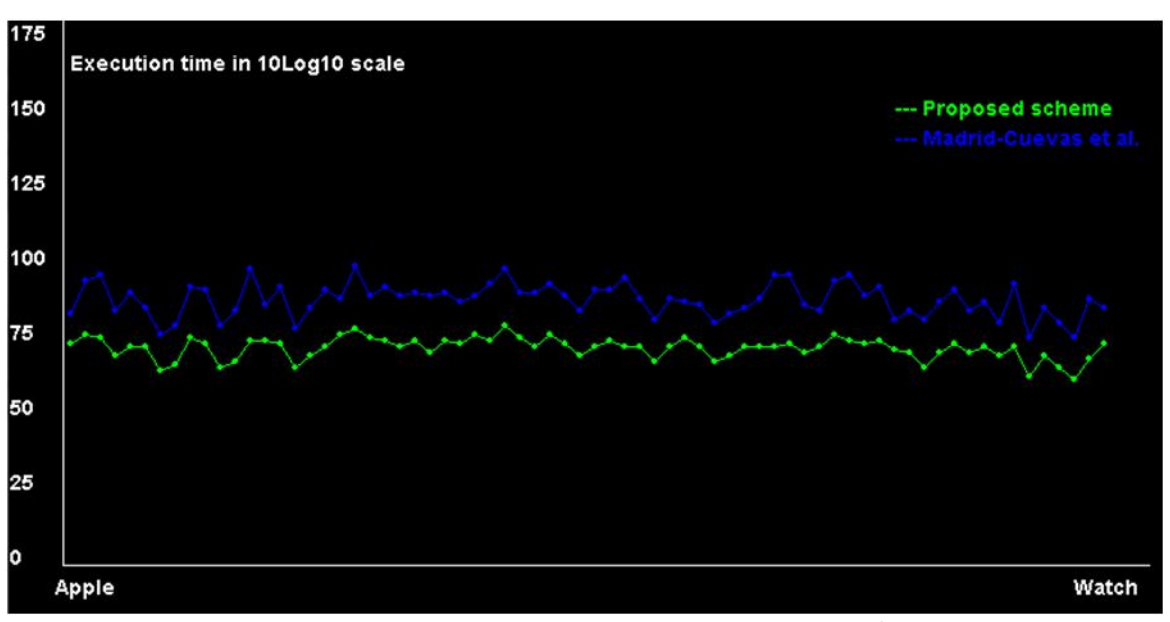

**FIGURE 6 The execution time in 10Log10 scale for the proposed scheme (green line diagram) and the Madrid-Cuevas et al. scheme (blue line diagram)**

The horizontal axis (Figure 6) shows the digital curve from the MPEG7 dataset, and the vertical axis shows the execution time on a 10Log10 scale. The blue line diagram shows the execution time for the Madrid-Cuevas et al. scheme, and the green line diagram is the plot of execution time for the proposed scheme. As seen from the graph, the proposed scheme has a shorter execution time than the Madrid-Cuevas et al. scheme. An analysis of the execution time of the two schemes shows that the Madrid-Cuevas et al. scheme is on average slower than the proposed scheme by a factor of 1.22 in 10Log10 scale.

The proposed scheme is compared with the Madrid-Cuevas et al. scheme with respect to its robustness under geometric transformations such as rotation and scaling. The compactness, defined by the quotient of the area and the square of the perimeter of an approximating polygon, is considered here to measure the similarity among different polygonal approximations under geometric transformation. The curves from the MPEG7 dataset used as test curves in the earlier part of the experiments were subjected to rotation by ten to eighty degree with a step size of ten and the transformed integer coordinates were used to find their polygonal approximations. The MPEG7 dataset curves were also subjected to scaling by scale factors of less than unity and greater than unity. Duplicate coordinates generated by scaling by a factor of less than unity were deleted. The gaps produced between successive points of a curve because of scaling by a factor greater than unity were filled using Bresenham's points generated by Bresenham's digital differential analyzer. The scale factors that are less than unity start at 20 percent and vary up to 80 percent with a step size of 20. The scale factors that are greater than unity start with a factor of 120 percent and vary up to 200 percent with a step size of 20. To determine the extent to which the transformed polygons of a curve are similar with respect to compactness, a statistical measure of variability was used. Because there is a possibility for the average compactness to vary widely from each other for different curves, the coefficient of variation defined by the quotient of standard deviation and arithmetic mean expressed in percentage is used to measure the degree of similarity between the polygons of a curve and its transformed counterparts. The coefficient of variation of the compactness for each curve along with its transformed counterparts is computed for the proposed scheme and for the Madrid-Cuevas et al. scheme. Figure 7 shows the coefficient of variation (scaled) for the two schemes plotted for different curves, where it can be observed that the coefficient of variation for the proposed scheme is less than that for the Madrid-Cuevas et al. scheme, except for a few exceptional cases. This observation establishes that the approximations generated by the proposed scheme are more robust than the Madrid-Cuevas et al. scheme with respect to rotation and scaling.

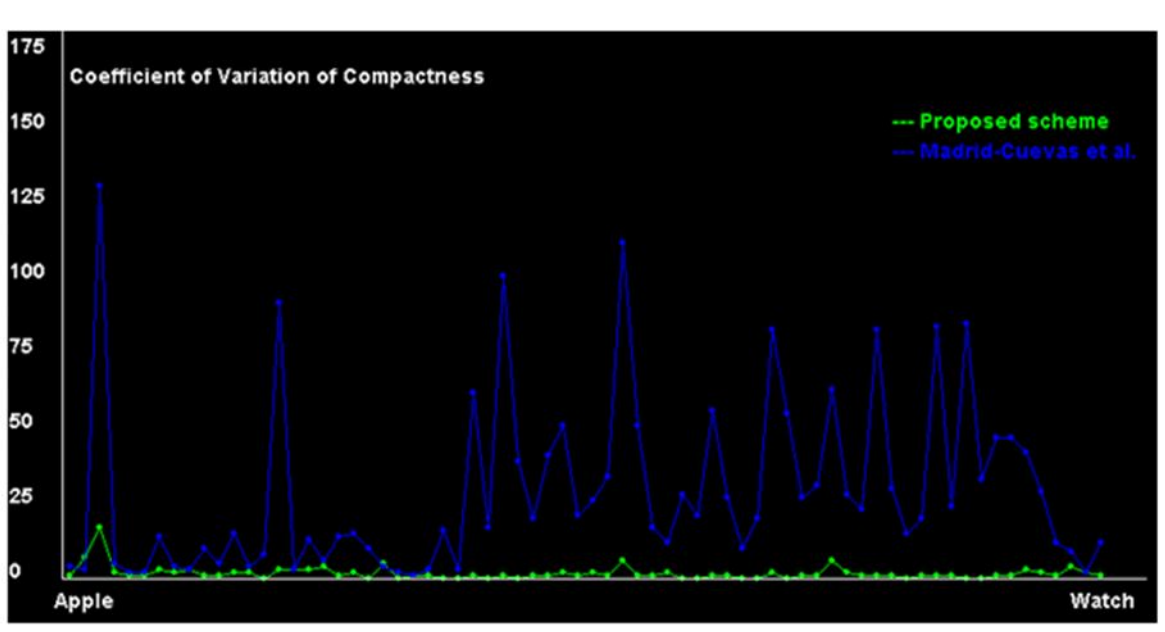

**FIGURE 7 Coefficient of variation of compactness for the proposed scheme and the Madrid-Cuevas et al. scheme**

## VI. CONCLUSION AND FUTURE RESEARCH

A fast and unsupervised scheme for polygonal approximation was developed. The scheme was compared with a competing scheme. It can be observed that the proposed scheme accurately reflects sharp vertices. The approximations generated by the proposed scheme are found to be competitive with the competing scheme with respect to Rosin's measure and aesthetic aspects, and are faster than the competing scheme. The proposed scheme was found to be more robust than the competing schemes with respect to geometric transformation.

Future research can be directed towards the search for a single-phase scheme (similar to near-convex decomposition in [24]), which should be fast, unsupervised, robust to geometric transformation and noise, and generate aesthetic approximation.